\documentclass[10pt]{bmc_article}

\usepackage{cite} 
\usepackage{url}  
\usepackage{ifthen}  
\usepackage{multicol}   
\usepackage[utf8]{inputenc} 
\usepackage{amsmath}
\usepackage{indentfirst}
\usepackage{graphicx}

\newboolean{publ}

\begin{document}


\title{\begin{center}Absorbing phenomena and escaping time for\\
 Muller's ratchet in adaptive landscape\end{center}}


\author{\begin{center}Shuyun Jiao$^{1,2}$%
         \email{ Shuyun Jiao- jiaoshuyun79@yahoo.com.cn}
         and Ping Ao\correspondingauthor$^{1,3}$%
  \email{Ping Ao\correspondingauthor -aoping@sjtu.edu.cn}\end{center}
}


\address{%
    \iid(1)Shanghai Center for Systems Biomedicine, State Key laboratory of Oncogenes and Related Genes, Shanghai Jiao Tong University, 200240, Shanghai, China\\
    \iid(2)Department of Mathematics, Xinyang Normal University, 464000, Xinyang, Henan, China \\
   \iid(3)Department of Physics, Shanghai Jiao Tong University, 200240, Shanghai, China
 }%

\maketitle


\setlength{\baselineskip}{20pt}
Publish information: 
BMC Systems Biology 2012, 6(Suppl 1):S10 ~~~~doi:10.1186/1752-0509-6-S1-S10

\begin{abstract}
\indent Background: The accumulation of deleterious mutations of a population directly  contributes to the fate  as to how long the population would exist, a process often described as Muller's ratchet with the absorbing phenomenon. The key to understand this absorbing phenomenon is to characterize the decaying time of the fittest class of the population. Adaptive landscape introduced by Wright, a re-emerging powerful concept in systems biology, is used as a tool to describe biological processes.  To our knowledge,  the dynamical behaviors for Muller's ratchet over the full parameter regimes are not studied from the point of the adaptive landscape.  And the characterization of the absorbing phenomenon is not yet quantitatively obtained without extraneous assumptions as well.

 \indent Methods: We describe how  Muller's ratchet can be mapped to the classical Wright-Fisher process in both discrete and continuous manners. Furthermore, we construct the adaptive landscape for the system analytically from the general diffusion equation. The constructed  adaptive landscape is independent of the existence and normalization of the stationary distribution.  We derive the formula of the single click time in finite and infinite potential barrier for all parameters regimes by mean first passage time.

\indent Results: We describe the dynamical behavior of the population exposed to Muller's ratchet in all parameters regimes by  adaptive landscape. The adaptive landscape has rich structures such as finite  and infinite potential, real  and imaginary fixed points. We give the formula about the single click time with finite and infinite potential. And we find the single click time increases with  selection rates and population size increasing, decreases with mutation rates increasing. These results provide a new understanding of infinite potential. We  analytically demonstrate the adaptive and unadaptive states for the whole parameters regimes. Interesting issues about the parameters regions with the imaginary fixed points is demonstrated.  Most importantly, we find that  the absorbing phenomenon  is characterized by the adaptive landscape and the single click time without any extraneous assumptions. These results suggest a graphical and quantitative framework to study the absorbing phenomenon.
\end{abstract}

\ifthenelse{\boolean{publ}}{\begin{multicols}{2}}{}


\section*{Background}
  Muller's ratchet proposed in 1964 is that the genome of an asexual population accumulates deleterious mutations in an irreversible manner. It is a mechanism that has been suggested as an explanation for the evolution of sex \cite{Maynard1978}.  For asexually reproducing population, without recombination, chromosomes are directly passed down to offsprings. As a consequence,  the deleterious mutations accumulate so that the fittest class loses. For sexually reproducing population, because of the existence of recombination between parental genomes,  a parent carrying high mutational loads can have offspring with fewer deleterious mutations. The high cost of sexual reproduction is thus offset by the benefits of inhibiting the ratchet \cite{Etheridge2009}.  Muller's ratchet has received growing attention recently. Most studies of Muller's ratchet are related to two issues. One is that without recombination, the genetic uniformity of the offspring leads to much lower genetic diversity, which is likely to make it more difficult to adapt \cite{Lampert2010}. So its adaptiveness arouses concern. The other is that population lacking genetic repair should decay with time, due to successive loss of the fittest individuals \cite{Maia2009}\cite{Barton2010}. So the fixation probability arouses concern. In addition, Muller's ratchet is relevant to some replicators \cite{Gordo2000a}\cite{Gordo2000b}, endosymbionts \cite{Moran1996}, and mitochondria \cite{Bergstrom1998}. In order to assess the relevance of Muller's ratchet, it is necessary to determine the rate (or the time) for the accumulation of deleterious mutations \cite{waxman2010}. It is widely recognized that the rate of deleterious mutations being much higher than that of either reverse or beneficial mutations results in a serious threat to the survival of populations at the molecular level \cite{Maia2009}.   Because models proposed must rest on the biological reality, which must be analyzed on their own without any injection of extraneous assumptions during the analysis \cite{ewens1979}.  Overall, it has been a long interest to develop a suitable and quantitative theory for the ratchet mechanism and the incidental absorbing phenomenon.\\
\indent Biologists  have suggested  \cite{felsenstein1974} \cite{waxman2010} that a quantitative framework is needed. The potential evolutionary importance of Muller's ratchet makes it desirable to carry out careful quantitative studies \cite{felsenstein1974}. And the incidental absorbing phenomenon is investigated quantitatively in broad literature.  The simplest and earliest mathematical model is
the pioneering work in  Ref.\cite{Haigh1978}. It described the same evolutionary process on the condition of deterministic mutation-selection
balance according to the Wright-Fisher dynamics. And it
indicated numerical evidence of relation between the total number of individuals and
the average time between clicks of the ratchet, but it did not focus on the absorbing phenomenon. It treated the pioneering model as a diffusion
approximation \cite{stephan1993}, and produced more accurate predictions over the
relatively slow regime. It noted that the increasing importance of selection coefficients
for the rate of the ratchet for increasing values of the total number of individuals. But it is represented as stochastic differential equations and did not get the predictions over all parameters regions.
It employed simulation
approaches to Muller's ratchet \cite{Gessler1995} and estimated how different between
the distribution of mutations within a population and a Poisson distribution. But it did not emphasize the  absorbing phenomenon.  In  Ref.\cite{Etheridge2009} it obtained diffusion
approximations for three different parameter regimes, depending on
the speed of the ratchet. The model shed new light on \cite{stephan1993}. But it mainly focused on the property of the solution for these stochastic differential equations.
 In Ref.\cite{waxman2010} it mapped Muller's ratchet to
Wright-Fisher process, and got the prediction of the rate of accumulation of deleterious mutations when parameters lie in the fast and slow regimes of the operation of the ratchet. But it put the constraints of Dirac function  on the boundary. Previous works mainly focused on the parameter regimes with  lower or  higher mutation rates. And models are represented as stochastic differential equation.  In Ref. \cite{Higgs1996} authors imagined the population evolved on an adaptive landscape, but they could not analytically construct it. It described discrete birth-death model and  its corresponding diffusion manner by the adaptive landscape \cite{zhou2011}. But it did not discuss the absorbing phenomenon.  The concept of
adaptive landscape is proposed by Sewall Wright to build intuition for the complex biological phenomena  \cite{Barton2010}.  In the present article,
inspired by  \cite{Higgs1996}\cite{waxman2010}, we model Muller's ratchet as a Wright-Fisher process,
analytically construct the adaptive landscape, where the non-normalizable stationary distribution occurs. Here the adaptive landscape is analytically quantified as a potential function  from the physical point of view \cite{Ao2005}. We give the adaptive and unadaptive states for the whole parameters region by the adaptive landscape. We give the formula for the single click time of Muller's ratchet in the face of infinite and finite potential.  In addition, we can handle the absorbing phenomenon without extraneous assumptions.\\
\indent The key concept in constructing the adaptive landscape is of potential function as a scalar function.
There is a long history of definition, interpretation,
and generalization of the potential. Such potential has also been applied to biological systems in various ways. The usefulness of a potential reemerges in the current
 study of dynamics of gene regulatory networks \cite{Ao2004A}, such as its application in genetic switch  \cite{Zhu2004}\cite{Zhu2004a}\cite{liang2010}\cite{cao2010}. The role of potential is the  same as that of adaptive landscape. In this article, we do not distinct them.\\
\indent We now make the obvious advance to Muller's ratchet. We analytically construct the adaptive landscape. We demonstrate the position and adaptiveness of fixed points. This makes the dynamical behaviors of the population to be investigated. In addition, we give the area with imaginary fixed points. This makes the explaining for the imaginary fixed points biologically possible.  Infinite potential barriers can be crossed over under some cases. We handle the absorbing phenomenon without any extraneous assumptions under the condition of diffusion approximation.   Inversely, we demonstrate the power of the adaptive landscape.
\section*{Methods}
\subsection*{Discrete model and absorbing boundary}
\indent  We consider here in population genetics an important and widely applied mechanism-- Muller's ratchet. It is the process by which the genome of an asexual population accumulates deleterious mutations in an irreversible manner \cite{muller1932}\cite{muller1964}. It  corresponds to the repeated irreversible loss of the fittest class of individuals because of the accumulation  of the deleterious mutations, the effective absence of beneficial mutations, without any  recombination, but with the random  drift \cite{muller1964}\cite{felsenstein1974}. Consider a population of haploid asexual individuals with discrete generations $t=0,1,2,\ldots$. The common point in a generation is regarded as an adult stage, after all selection has occurred and immediately prior to reproduction.  New mutations occur at reproduction and all mutations are assumed to deleteriously affect viability but have no effect on fertility. Supposed population size is fixed for each generation.  The viability of a newly born individual is taken to be determined solely by the alleles they carry. This allows us to divide the population into different classes with different genotypes. \\
\indent Here in one dimensional case, we consider one locus with two alleles (for example, $A$ and $a$), that is, there are two classes in the haploid asexual population, one class with allele $A$ while the other with allele $a$, supposed mutation from allele $A$ to $a$ is deleterious. We assume fixed population size of $N$, which means we  have $N$ alleles in all. We also assume that $N>1$. Generations are non-overlapping. The lifecycle of the individuals in the population is from adults to juveniles, during which we consider irreversible mutation, selection, and random drift. The frequency of the allele $A$ for generation $t$ is $\overline{p}$ while that of allele $a$ is $1-\overline{p}$.  Let $\mu$ be the probability that an offspring of an adult with allele $A$ is an individual with allele $a$, labeled by $M_{1,0}$, that is $M_{1,0}=\mu$. Analogously, $M_{0,0}=1-\mu$, $M_{0,1}=0$, $M_{1,1}=1$. The relative viability of individuals with allele $A$ is $\nu_{0}=1$ while that of individuals with allele $a$ is $\nu_{1}=1-\sigma$. Where $\sigma$ can be treated as an effective selection coefficient associated with deleterious mutations. So the values of parameters for $\mu$ and $\sigma$ are from $0$ to $1$. Then in generation $t+1$, when selection and deleterious mutation are active, the probability that the offspring of a parent with allele $A$ is chosen to be with allele $a$ is $\mu\overline{p}(1-\sigma)$, the probability that the offspring with allele $A$ is $(1-\mu) \overline{p}$, the probability that the offspring of a parent with allele $a$ is still with allele $a$ is $(1-\sigma)(1- \overline{p})$. So the frequency of allele $A$ in generation $t+1$ is
\begin{equation}
\label{1}
\overline{p}(t+1)=\frac{(1-\mu)\overline{p}(t)}{1-\sigma+\sigma(1-\mu)\overline{p}(t)}.
\end{equation}
 Eq.(1) describes the deterministic process that ignores random drift. Under the mutation-selection balance, the fixed points are
\begin{eqnarray}
\overline{p}_{1}(t=\infty)&=&0,\\
\overline{p}_{2}(t=\infty)&=&\frac{\sigma-\mu}{\sigma(1-\mu)}.
\end{eqnarray}
 This means the population ultimately arrives at the state with allele frequency $\overline{p}_{1}$ or $\overline{p}_{2}$ and no transition between the two states occur. It is evident that selection rates $\sigma$ is greater than mutation rates $\mu$. But  populations always evolve randomly. Since each individual is assigned a parent independently, if by generation $t$ the average value of allele frequency is $p_{n,t}$, we have that the transition probability from  $p_{n,t}$ to $p_{n,t+1}$ by generation $t+1$  is 
\begin{eqnarray}
W_{nm}=\frac{N!}{m!(N-m)!}(\overline{p}_{n,t})^{m}(1-\overline{p}_{n,t})^{N-m}.~ ~ ~n,m=0,1,2,\ldots,N.
\end{eqnarray}
The matrix of transition probabilities  is $\bf W$ which is composed of elements $W_{nm}$. The dynamical rule can be written as the matrix Eq.(5)
 \begin{equation}
 \label{2}
 {\bf P}(t+1)={\bf WP}(t).
 \end{equation}
Where ${\bf P}(t)$ represents the probability distribution of allele $A$. It is  composed of $N+1$ elements $\widetilde{p_{n}}(t)$, $n=0,1,\ldots,N$, where $\widetilde{p_{n}}(t)$ denotes the probability that allele $A$ has the frequency $n/N$ in generation $t$ in the presence of deleterious mutation, selection and random drift. The matrix of transition probabilities  is
\begin{eqnarray}
{\bf W}=\left(\begin{array}{cc}
1&\bf{v^{T}}\\
\bf{0}& \bf{w}
\end{array}\right).
\end{eqnarray}
Here $\bf{v^{T}}$ means the transpose of vector $\bf{v}$. Then Eq.(5) can be expressed as the following
\begin{eqnarray}
\left(\begin{array}{c}
\widetilde{p_{0}}(t+1)\\
{\bf p}(t+1)
\end{array}\right)
=\left(\begin{array}{cc}
1&\bf{v^{T}}\\
\bf{0}& \bf{w}
\end{array}\right)
\left(\begin{array}{c}
\widetilde{p_{0}}(t)\\
{\bf p}(t)
\end{array}\right).
\end{eqnarray}
Where $\bf{0}$ is a column vector with $N$ vanishing components, and $\bf{v}$ is a column vector with $N$ non-zero elements given by $v_{m}=W_{0m}$, where $m$ is from $1$ to $N$. The column vector $\bf{w}$ is an $N\times N$ matrix with elements $w_{nm}=W_{nm}$, where $n$ and $m$ run from $1$ to $N$. By spectrum decomposition we can derive the maximum eigenvalue for $\bf w$, the corresponding eigenvector is the quasi-stationary probability density of allele $A$. The following is its derivation process.
\begin{equation}
{\bf p}(t+1)={\bf wp}(t),
\end{equation}
\begin{equation}
\widetilde{p_{0}}(t+1)=\widetilde{p_{0}}(t)+{\bf v^{T}p}(t),
\end{equation}
Eq.(8) has the solution
\begin{equation}
{\bf p}(t)={\bf w}^{t}{\bf p}(0).
\end{equation}
Because the leading large time behavior is determined by the eigenvalue of matrix $\bf{w}$. The issue is transformed to solve the leading large eigenvalue of $\bf{w}$ and corresponding eigenvector.
But
\begin{eqnarray}
\widetilde{p_{0}}(t)&=&1-\sum_{n=1}^{N}\widetilde{p_{n}}(t)\nonumber \\
&=&1-{\bf I^{T}p}(t)\nonumber \\
&=&1-{\bf I^{T}w}^{t} {\bf p}(0).
\end{eqnarray}
Where $\bf{I^{T}}$$=(1,\ldots,1)$, there are $N$ elements $1$ of this vector. If we denote the leading large eigenvalue of $\bf{w}$ is $\lambda_{1}$,
\begin{eqnarray}
1-\widetilde{p_{0}}(t)&=&{\bf I^{T}w}^{t} {\bf p}(0)\nonumber\\
&\propto&\lambda_{1}^{t}.
\end{eqnarray}
We can describe Eq.(8) as
\begin{equation}
{\bf p}(t)\approx [1-\widetilde{p_{0}}(t)]{\bf q}(t),
\end{equation}
where
\begin{equation}
{\bf q}(t)=\frac{{\bf w}^{t}{\bf p}(0)}{{\bf I^{T}w^{T}p}(0)}.
\end{equation}
Then the probability density is derived by
\begin{equation}
{\bf q}=\lim_{t\rightarrow \infty}\frac{{\bf w}^{t}{\bf p}(0)}{{\bf I^{T}w^{T}p}(0)}.
\end{equation}
Where $\lambda_{1}$ is determined by $1-\widetilde{p_{0}}(t)$. But $1-\widetilde{p_{0}}(t+1)\approx (1-{\bf v^{T}q})[1-\widetilde{p_{0}}(t)]$. In the end we get $\lambda_{1}=1-{\bf v^{T}q}$. \\
\indent Populations evolution is a natural and random process. We model the process as the discrete form. The boundary of the discrete model is determined by its transition matrix. Generally, The transition probabilities from  generation $t$ to generation $t+1$ for the allele frequency are expressed as
\begin{equation}
\frac{N!}{m!(N-m)!}\left(\frac{(1-\mu)p_{n,t-1}}{1-\sigma+\sigma(1-\mu)p_{n,t-1}}\right)^{m}\left(1-\frac{(1-\mu)p_{n,t-1}}{1-\sigma+\sigma(1-\mu)p_{n,t-1}}\right)^{N-m}.
\end{equation}
 From the expression of transition probability, it can be seen that the transition probabilities are zero for any frequency $p_{n,t-1}$ under the condition of parameter $\sigma=1$. It means the population stays at its initial state. In addition, the transition probabilities from the boundary  $0$ to its next are $(1,0,\ldots,0)^{T}$. This means boundary $0$ can not output any probability flow to its next, it only absorbs probability from next. We call absorbing phenomenon occurring at the boundary $0$.
\subsection*{Continuous model and adaptive landscape}
\subsubsection*{Diffusion approximation}
\indent Here we briefly outline the diffusion approximation from the discrete to continuous models.   At generation $t$ the frequency of allele $A$ is $i/N$,  after evolutionary force, at generation $t+1$ the allele frequency becomes $j/N$. Here $\delta t=1$, the probability that allele frequency becomes $j/N$ is
\begin{equation}
\widetilde{p}_{j}(t+1)=\sum_{i=0}^{N}W_{ij}\widetilde{p}_{i}(t).
\end{equation}
Where $W_{ij}$ is the transition probability. Diffusion approximation is a description of the process, valid when $N$ is large, where the allele frequency $n/N$ are replaced by real number $x$, $0\leq x \leq 1$. Given that $A$ starts out at gene frequency $x_{0}$. Let  $\widetilde{p}(x-\delta x,t)$ be the probability for allele frequency $A$ after $t$ generations. And $\widetilde{p}(x,t+1)$ be the probability of allele $A$ after $t+1$ generations, then
\begin{equation}
\widetilde{p}(x,t+\delta t)=\sum_{\delta x}W(x,t+\delta t|x-\delta x,t)\widetilde{p}(x-\delta x,t).
\end{equation}
Letting $\widetilde{p}(x,t)=\rho(x,t)/N$, among this $\rho(x,t)$ is the probability density. Define $M(x)$ as the probability that $x$ increases by systematic force that include mutation and selection. And define $V(x)$ as the probability  that $x$ changes because of random drift, either decreasing by amount $\delta x$ with the probability $V(x)/2$ or increasing by the amount $\delta x$ with the probability $V(x)/2$.
\begin{eqnarray}
M(x,t)&=&\lim_{\delta t\rightarrow 0}\frac{1}{\delta t}\int_{0}^{1}\delta x W(x,t+\delta t|x-\delta x,t)d(\delta x),\\
V(x,t)&=&\lim_{\delta t\rightarrow 0}\frac{1}{\delta t}\int_{0}^{1}(\delta x)^{2}W(x,t+\delta t|x-\delta x,t)d(\delta x).
\end{eqnarray}
In any time interval $\delta t$, the probability that $x$ remains at $x$ is $1-M(x)-V(x)$. The changes in states are only to $\delta x$ or $-\delta x$. So $\delta x$ is $0$, positive or negative.
\begin{equation}
\rho(x,t+\delta t)=(1-M(x)-V(x))\rho(x,t)+M(x-\delta x)\rho(x-\delta x,t)+\frac{1}{2}V(x+\delta x)\rho(x+\delta x,t)+\frac{1}{2}V(x-\delta x)\rho(x-\delta x,t),
\end{equation}
and we can treat Eq.(21) as the following
\begin{eqnarray}
\rho(x,t+\delta t)-\rho(x,t)&=&-[M(x)\rho(x,t)-M(x-\delta x)\rho(x-\delta x,t)]\nonumber\\
&&\quad +\frac{1}{2}[V(x+\delta x)\rho(x+\delta x,t)
 -V(x)\rho(x,t)\nonumber\\
&&-V(x)\rho(x,t)+V(x-\delta x)\rho(x-\delta x,t)],
\end{eqnarray}
that is
\begin{equation}
\frac{\rho(x,t+\delta t)-\rho(x,t)}{\delta t}=-\frac{\delta[M(x)\rho(x,t)]}{\delta x}+\frac{1}{2}\frac{\delta (\delta[V(x)\rho(x,t)])}{\delta(\delta x)}.
\end{equation}
Because the function about the change of allele frequency in one generation is continuous and smooth enough, under the condition  that $|\widetilde{p}_{j,t+1}-\widetilde{p}_{i,t}|>1/N$ is of order of magnitude small than $\mathcal{O}(1)$ in one generation. Put it differently, the change of allele frequency is not more than $1/N$, and the probability density is smooth enough during the time scale of one generation. So we represent the process  as  the following approximated diffusion equation
\begin{equation}
\frac{\partial}{\partial{t}}\rho(x,t)=-\frac{\partial}{\partial{x}}[M(x)\rho(x,t)]+\frac{1}{2}\frac{\partial^{2}}{\partial{x}^{2}}[V(x)\rho(x,t)],
\end{equation}
and according to the definition of $M(x)$ and $V(x)$, the explicit expressions of them are
\begin{eqnarray}
M(x)&=&\frac{(1-\mu)x}{1-\sigma+\sigma(1-\mu)x}-x \nonumber\\
&=&\frac{x[(\sigma-\mu)-\sigma(1-\mu)x]}{1-\sigma+\sigma(1-\mu)x},\\
V(x)&=&\frac{x(1-x)}{N}.
\end{eqnarray}
Among this $M(x)$ is the symbol for the change in allele frequency \cite{Kimura1964}\cite{ewens1979} that occurs in one generation due to systematic force. The function $V(x)$ is the variance in allele frequency after one generation of binomial sampling of $N$ alleles \cite{John1998}.
\subsubsection*{Adaptive landscape}
\indent Under the general diffusion approximation, frequency $p_{n,t}$ is treated as continuous quantities $x$, and this leads to the distribution of the frequency for the  allele $A$ being the probability density. Let $\rho(x,t)$ be the probability density of the
frequency for the allele $A$ being $x$ at time $t$. The diffusion
process can be expressed by the following symmetric equation
\begin{equation}
\label{2}
\partial_{t} \rho(x,t) = \partial_{x} [\epsilon D(x)\partial_{x}-f(x)] \rho(x,t)
\end{equation}
with
 \begin{eqnarray}
\label{3}
    f(x) &=&M(x)-\epsilon D'(x),\nonumber \\
    2\epsilon D(x) &=&V(x).
\end{eqnarray}
With a prime denoting differentiation of a function with  respect to its argument such as $D'(x)\equiv \partial_{x} D(x)$. Where $M(x)$ and $V(x)$ is from  Eqs.(25) and (26) respectively. Adaptive landscape is directly given when we consider natural boundary as Feller's classification. It is
\begin{equation}
\Phi(x)=\int \frac{f(x)}{D(x)}dx.
\end{equation}
\indent The symmetric Eq.(27) has two advantages. On the one hand, the adaptive landscape is directly read out when the detailed balance is satisfied. On the other hand, the constructive method is dynamical, independent of existence and normalization of stationary distribution.  We call $f(x)$ directional transition rate, integrating the effects of $M(x)$ and the derivative of $V(x)$. Directional transition rate can give equilibrium states when it is linear form. \\
\indent When the process lies at stationary state, the probability flux of the system is zero, and
probability flux flows in $x\in[0,1]$. In general, the stationary distribution for the diffusion approximation satisfying natural boundary condition is given by
\begin{eqnarray*}
\label{5}
\rho(x,t=\infty)&=& \dfrac{1}{Z}\exp \left(\dfrac{\Phi(x)}{\epsilon}\right).\\
Z&=&\int_{-\infty}^{+\infty} \exp\left(\dfrac{\Phi(x)}{\epsilon}\right)dx.
\end{eqnarray*}
It has the form of Boltzmman-Gibbs distribution \cite{Ao2008A}, so the scalar function $\Phi(x)$ naturally acquires the meaning of potential energy \cite{Ao2004A}. The value of $Z$ determines the normalization of $\rho(x,t=\infty)$ from the perspective of probability,
  and the finite value of $Z$ manifests the normalization of $\rho(x,t=\infty)$. The stationary distribution is not true in the face of infinite $Z$. It demonstrates absorbing phenomenon occurs at the boundary. Together with the flux at the boundary, the true stationary distribution could be got. The constant $\epsilon$ holds the same position as temperature of Boltzmman-Gibbs distribution in statistical mechanics. But it does not hold the nature of temperature in Boltzmman-Gibbs distribution.\\
\indent We are interested in the dynamical property of adaptive landscape,  so we treat $\Phi$ and $\Phi/ \epsilon$  no difference in this respect, that is, for convenience we can take $\epsilon=1$ of $\epsilon D(x)$. So according to Eq.(29), we have adaptive landscape as the following
\begin{eqnarray}
\label{7}
\Phi(x)&=&\frac{2N \mu(1-\sigma)}{1-\sigma \mu}\ln(1-x)-\ln x(1-x)   \nonumber \\
\quad &&+\frac{2N(1-\mu)}{1-\sigma \mu}\ln(1-\sigma+x\sigma(1-\mu)).
\end{eqnarray}
From the expression  of adaptive landscape
$\Phi(x)$, we may find there are two singular points $0$ and $1$ of adaptive landscape, characterized by infinite value, infinity means adaptiveness or unadaptiveness  of the system. Here the adaptive landscape is composed of three terms. The first term and the third term quantify the effect of  selection and irreversible mutation respectively, the second term quantifies the effect of random drift. \\
\indent The stationary distribution can  be expressed as
\begin{eqnarray}
  \rho(x,t=\infty)& \propto&  \exp \left(\frac{\sigma \mu-1-2N \mu(\sigma-1)}{1-\sigma \mu}\ln(1-x)\right. \nonumber \\
&& \quad \left .-\ln x + \frac{2N(1-\mu)}{1-\sigma \mu}\ln(1-\sigma+x\sigma(1-\mu))\right).
\end{eqnarray}
\section*{Results}
\subsection*{Fixed points and their adaptiveness}
\indent To understand the mechanism of Muller's ratchet, a full characterization of dynamical process is a prerequisite for obtaining  more accurate decaying time. Here we study the dynamical behaviors by investigating the position and adaptiveness of all fixed points. We further derive the parameter regions for all possible cases. \\
\indent According to general analysis of a dynamical system, letting
\begin{equation}
\label{18}
\Phi'(x)=0,
\end{equation}
we get
\begin{equation}
2\sigma(1-\mu)(N-1)x^2+(2N(\mu-\sigma)+3\sigma-\sigma\mu-2)x+(1-\sigma)=0.
\end{equation}
We solved the Eq.(32) and found two fixed points. If we denote
\begin{eqnarray}
\alpha&=&2-3\sigma+\sigma\mu+2N\sigma-2N\mu,\\
\beta&=& 8\sigma(1-\mu)(N-1)(1-\sigma).
\end{eqnarray}
They are
\begin{equation}
\label{19}
x_{1,2}=\frac{\alpha\mp \sqrt{\alpha^{2}-\beta}}{4\sigma(1-\mu)(N-1)}.
\end{equation}
\indent For two singular points $x=0,1$, if $x\rightarrow 1$, and $\sigma\in(\mu,(2N\mu-1)/(2N\mu-\mu))$, $\Phi(x)\rightarrow -\infty$. So the population is unadaptive at $x=1$. When $x\rightarrow 1$, and $\sigma\in((2N\mu-1/(2N\mu-\mu),1)$,  $\Phi(x)\rightarrow +\infty$. So the population is adaptive at $x=1$. For $x\rightarrow 0$, $\Phi(x)\rightarrow +\infty$ in almost parameters regimes except $\sigma=1$. So the population is always adaptive at $x=0$.  When $\sigma=1$, the viability of the suboptimal class is zero, so populations stay at the initial state, the corresponding  minimum of adaptive landscape demonstrates  the state with allele frequency $x=0$.\\
\indent Here we address dynamical behavior by the positions of two real inequivalent fixed points $x_{1}<x_{2}$ first. \\
\indent I) We find two different real fixed points in the regimes of $\mu\in(0,2/(2N-1+2\sqrt{N(N-1)}))$ and $\sigma\in(\mu,1)$; in the regimes of $\mu\in(2/(2N-1+2\sqrt{N(N-1)}),1)$ and $\sigma\in((2+2\mu-10N\mu+4N^{2}\mu+2N\mu^{2}+4(1-\mu) \sqrt{N(N-1)((2N-1)\mu-1)})/(\mu-2N+1)^{2},1)$ except the regime of  $\mu\in((2N-1)/4N(N-1), 1)$ and $\sigma=(2N\mu-1)/(2N\mu-\mu)$.\\
\indent We discuss the position between them and the boundary points $0$, $1$ and adaptiveness of them in the following.\\
\indent i) $1<x_{1}<x_{2}$\\
In the regimes of $\mu\in(1/(2N-1),2/(2N-1+2\sqrt{N(N-1)})]$ and $\sigma\in(\mu,(2N\mu-1)/(2N\mu-\mu))$; in the regimes of $\mu\in(2/(2N-1+2\sqrt{N(N-1)}),(2N-1)/4N(N-1))$ and $\sigma\in((2+2\mu-10N\mu+4N^{2}\mu+2N\mu^{2}+4(1-\mu)\sqrt{N(N-1)((2N-1)\mu-1)})/(\mu-2N+1)^{2},(2N\mu-1)/(2N\mu-\mu))$, the fixed points satisfy $1<x_{1}<x_{2}$. At the same time the singular point $x=1$ is unadaptive. There is one adaptive state with allele frequency $x=0$ in the system. Populations tend to evolve to the adaptive state.\\
\indent ii) $1=x_{1}<x_{2}$\\
 In the regions of  $\mu\in(1/(2N-1),(2N-1)/4N(N-1))$ and $\sigma=(2N\mu-1)/(2N\mu-\mu)$, the two fixed points satisfy $x_{1}=1$, $1<x_{2}$. The state with allele frequency $x=1$ is unadaptive. There is one adaptive state with allele frequency $x=0$ in the system. \\
\indent iii) $0<x_{1}<1<x_{2}$ \\
In the regimes of  $\mu\in(0,1/(2N-1))$ and $\sigma\in(\mu,1)$; in the regimes of $\mu\in(1/(2N-1),1)$ and $\sigma\in((2N\mu-1)/(2N\mu-\mu),1)$, the fixed points satisfy $0<x_{1}<1$, $1<x_{2}$.  There is only one unadaptive state with allele frequency $x=x_{1}$ in the system, and two adaptive states with allele frequency $x=1$ and $x=0$ occur in the system. Populations tend to evolve to the adaptive state dependent on the position of the initial state. If the initial state with allele frequency is greater than $x_{1}$, populations tend to evolve to the adaptive state with allele frequency $x=1$. \\
\indent iv) $0=x_{1}<1<x_{2}$\\
In the regime of  $\mu\in(0,1)$ and $\sigma=1$, the fixed points satisfy $x_{1}=0$, $1<x_{2}$. When selection rate $\sigma=1$, the process lies at the initial state because for this case, the viability of the sub-fittest class is zero.  \\
\indent v) $0=x_{1}<x_{2}<1$\\
The case $0=x_{1}<x_{2}<1$ is impossible. For $x_{1}=0$, that is $\sigma=1$, at the same time $x_{2}$ must be greater than one.\\
\indent vi) $0<x_{1}<x_{2}<1$ \\
In the regimes of $\mu\in((2N-1)/4N(N-1),(2N-1)/(4N-3))$ and $\sigma\in((2+2\mu-10N\mu+4N^{2}\mu+2N\mu^{2}+4(1-\mu)\sqrt{N(N-1)((2N-1)\mu-1)})/(\mu-2N+1)^{2},(2N\mu-1)/(2N\mu-\mu))$; in the regimes  $\mu\in((2N-1)/(4N-3),1)$ and  $\sigma\in((2+2\mu-10N\mu+4N^{2}\mu+2N\mu^{2}+4(1-\mu)\sqrt{N(N-1)((2N-1)\mu-1)})/(\mu-2N+1)^{2},(2N\mu-1)/(2N\mu-\mu))$, the fixed points satisfy $0<x_{1}<x_{2}<1$. The state with allele frequency  $x_{1}$ is unadaptive while that with allele frequency $x_{2}$ is adaptive. There are two adaptive states with allele frequency $x=0$ and $x=x_{2}$ and two unadaptive states with allele frequency $x=1$ and $x=x_{1}$ in the system. Populations evolve to which adaptive states dependent on the initial position. \\
\indent vii) $x_{1}<0$ or $x_{2}<0$\\
The case $x_{1}<0$ is impossible, and the case $x_{2}<0$ is impossible.\\
\indent II) Then we discuss the case of two equivalent real fixed points $x_{2}= x_{1}$.\\
\indent  In the regimes of $\mu\in(2/(2N-1+2\sqrt{N(N-1)}),1)$ and $\sigma= (2+2\mu-10N\mu+4N^{2}\mu+2N\mu^{2}+4(1-\mu) \sqrt{N(N-1)((2N-1)\mu-1)})/(\mu-2N+1)^{2}$, we find  two same fixed points
\begin{equation}
\label{21}
x_{1,2}=\dfrac{\alpha}{4\sigma(1-\mu)(N-1)}.
\end{equation}
\indent i) $1<x_{1,2}$ \\
In the regimes of $\mu\in(2/(2N-1+2\sqrt{N(N-1)}),(2N-1)/4N(N-1))$ and $\sigma=(2+2\mu-10N\mu+4N^{2}\mu+2N\mu^{2}+4(1-\mu) \sqrt{N(N-1)((2N-1)\mu-1)})/(\mu-2N+1)^{2}$, there are two same fixed points satisfy $1<x_{1,2}$, and they are unadaptive. There is one adaptive state with allele frequency $x=0$ in the process.\\
\indent ii) $1=x_{1,2}$\\
At the two points of  $((2N-1)/4N(N-1),2N/(2N-1)^{2})$ and $((2N-1)/(4N-3),(4(N-1)(3-6N+4N^{2})+8(N-1)(4N-3)\sqrt{N(N-1)/(4N-3)})/(2N-1)^{2})$, there are two same fixed points satisfy $x_{1,2}=1$, and they are unadaptive. There is one adaptive state with allele frequency $x=0$ in the process. \\
\indent iii) $0<x_{1,2}<1$ \\
In the regime of $\mu \in((2N-1)/4N(N-1),(2N-1)/(4N-3))$, $\sigma=(2+2\mu-10N\mu+4N^{2}\mu+2N\mu^{2}+4(1-\mu) \sqrt{N(N-1)((2N-1)\mu-1)})/(\mu-2N+1)^{2}$; in the regimes of $\mu\in((2N-1)/(4N-3),1)$ and  $\sigma=(2+2\mu-10N\mu+4N^{2}\mu+2N\mu^{2}+4(1-\mu) \sqrt{N(N-1)((2N-1)\mu-1)})/(\mu-2N+1)^{2}$,  there are two same fixed points satisfy $0<x_{1,2}<1$, and they are unadaptive. There is one adaptive state with allele frequency $x=0$ in the process.\\
\indent III) Finally we consider two imaginary fixed points $|x_{1}|=|x_{2}|$. Where the $|.|$ denotes the length for an imaginary points.\\
\indent In the regime of $\mu\in(2/(2N-1+2\sqrt{N(N-1)}),1)$ and $\sigma\in(\mu,(2+2\mu-10N\mu+4N^{2}\mu+2N\mu^{2}+4(1-\mu) \sqrt{N(N-1)((2N-1)\mu-1)})/(\mu-2N+1)^{2})$, there are two imaginary fixed points in the system. There is only one adaptive state with allele frequency $x=0$. Populations always evolve to the adaptive state.\\
\indent Especially $\Phi'(x)=0$ is a linear equation, the fixed point is read out from the expression of $f(x)$. We can measure the adaptiveness by the value of adaptive landscape.  The bigger the value of adaptive landscape is, the more adaptive  the corresponding  state would be. The corresponding area of the fixed points in parameters plane is the Fig.1.

\subsection*{Irreversible mutation, selection and random drift balance}
 \indent Concretely we divide mutation rates into three regimes. One is with mutation rates $\mu \in (0,1/(2N-1))$ and selection rates $\sigma \in(\mu,1)$. Another is  mutation  rates $\mu \in(1/(2N-1),(2N-1)/(4N(N-1))]$ and selection rates $\sigma \in(1/(2N-1),(2N\mu-1)/(2N\mu-\mu))$. Another is with mutation rates $\mu \in((2N-1)/4N(N-1),1)$ and selection rates $\sigma \in(2N/(2N-1)^{2},(2N\mu-1)/(2N\mu-\mu))$. The first parameter regimes, a part of regions of I iii), corresponds to the case that mutation rates lie in the lower regime. And the adaptive landscape is U-shape. The  second parameters regime,  same of I i),  is the middle regime. Adaptive landscape with these parameters demonstrates the monotonic decreasing. The third parameters region, same of I vi),  demonstrates the  clicking process.  The adaptive landscape of full parameter regimes  is visualized as Fig.2. From the expression and visualization of adaptive landscape
$\Phi(x)$, we may find there are two singular points $0$ and $1$ of
adaptive landscape, characterized by infinity values in Fig.2. Singularity from the maximum of adaptive landscape indicates the
population being adaptive while singularity
from the minimum of adaptive landscape indicates the population  being unadaptive.\\

\indent Fig.2 demonstrates the whole process of the population evolution including the forming and losing the fittest class except $\sigma=1$ demonstrated by red line in the middle graphs. Because the viability of the class with allele $a$ is not zero and deleterious mutation is arbitrary. The adaptive state under the condition of the parameter $\sigma=1$ only means the initial states.  With increasing selection rates the fittest class $A$  forms quickly while with increasing mutation rates the fittest class $A$ loses. In the lower mutation rates regime green line describes the population is likely to move to the fittest class with increasing selection rates, the process is dominated by selection. Blue line and red line manifest the losing process of allele $A$. Because the mutation rates are lower, selection rates dependent of mutation rates are lower, these factors result in the change of fittest class is not easy. In the end there are two adaptive states demonstrated by green line  in the process. In the high mutation rates regime, blue line  describes the population is likely to move to the fittest class so
that the population exists in the form of coexistence of $A$ and $a$.  Because deleterious mutation rates are
higher, as a consequence allele $a$ occurs. Selection rates
dependent of mutation rates tends to survive allele $A$. But the evolutionary process is dominated by the irreversible mutations, the fittest class $A$ loses. There are two adaptive states in the process under the balance of irreversible mutation and selection.  In the middle graph section, mutation rates and selection rates lie in the transition regime. It demonstrates the losing process demonstrated by red line. So we can draw the conclusion that the click process occurs when there are two stable states in the process.
\subsection*{Characterization of the single click time}
\indent We visualize the adaptive landscape, then one may wonder about how the population moves from one peak to another and how long it might be to move from one maximum to another. The process was first visualized by Wright in  1932. In addition, the problem of transition from metastable states is ubiquitous in almost all scientific areas.  Most of  previous works  encounter  finite potential barriers from the physical point of view.
  Interesting issue here is that we touch upon  infinite potential barriers under the circumstance of  well defined two adaptive states. Then we manifest the derivation of a single click time. The time of a click of the ratchet is recognized as the random time of loss of the fittest class  \cite{waxman2010}. \\
\indent The single click time  is well defined when there are two fittest classes in the process. It means the interval time between extinction of the two fittest classes. The corresponding processes are that there are two well-defined adaptive states in the system. Corresponding graphs of adaptive landscape is Fig.3.\\
\indent To evaluate the single click time and show the further power
of adaptive landscape, in the following we will demonstrate how the
single click time from one adaptive state to another is derived in this
framework. After straightforward calculation, backward Fokker- Planck equation
corresponding to Eq.(27) can be expressed with the property of time homogeneous in the following form \cite{Kampen1992}\cite{Bokse2003}
\begin{equation}
\label{21}
\partial_{t}\rho(x,t)=(f(x)+\epsilon D'(x))\partial_{x}\rho(x,t)+\epsilon D(x)\partial_{x}^2\rho(x,t).
\end{equation}
General single click time dependent on initial Dirac function satisfies
\begin{equation}
\label{22}
(f(x)+\epsilon D'(x))\partial_{x} T(x)+\epsilon D(x)\partial_{x}^2 T(x)=-1.
\end{equation}
With
\begin{eqnarray}
T'(1)=0,\nonumber\\
T(0)=0.
\end{eqnarray}
Above treatment is valid. Because populations evolution is according to Muller's ratchet, that is in the presence of deleterious mutation, without any recombination, but with selection and random drift. And the  model in discrete manner demonstrates the transition probabilities are $0$ from the boundary $x=0$ to its next. So the boundary $x=0$  only absorbs flux from its next, the boundary is absorbing. The probabilities from boundary $x=1$ is not zero because $\mu\neq 0$ and $\mu,\sigma\neq 1$. And the population can not be out of the boundary $x=1$. So the boundary $x=1$ is reflecting.
The general solution corresponding to Eq.(39) is
\begin{equation}
\label{25}
T(x)=\int_x^0 \frac{1}{\epsilon D(y)} \exp (-\Phi(y))dy\int_{1} ^ y \exp(\Phi(z))dz,
\end{equation}
here $\Phi(x)=\int ^x (f(x')/ D(x'))dx' (\epsilon=1)$. \\
\indent Here the evolutionary process  occurs when
$x\in[0,1]$.  We are more interested in the transition time
between the two adaptive states $x=0$ and $x=1$. In the process, there
are two important states $x^{\ast}$, $x_{0}^{\ast}$.   Interval
$(0,1)$ contains a potential well at $x^{\ast}$ and a potential barrier at $x_{0}^{\ast}$.
 The single click time is composed of two elements, one denotes forming process of fittest class, the other describes losing process of fittest class. In general, the time spent on forming process is much smaller than that spent on losing process. So the transition time approximates to the time spent on losing process. Because we assume that near $x_{0}^{\ast}$ we can write
\begin{eqnarray}
\Phi(x)\approx \Phi(x_{0}^{\ast})-\dfrac{1}{2}(\dfrac{x-x_{0}^{\ast}}{\alpha'})^{2}.
\end{eqnarray}
and near $x^{\ast}$
\begin{eqnarray}
\Phi(x)\approx \Phi(x^{\ast})+\dfrac{1}{2}(\dfrac{x-x^{\ast}}{\beta'})^{2}.
\end{eqnarray}
At the same time, if the central maximum of $\Phi(x)$ is large compared with $1/N$, then $\exp(\Phi(z))$ is sharply peaked at $x_{0}^{\ast}$, while $\exp(-\Phi(y))/D(y)$ is very small near $y=x^{\ast}$.  Eq.(41) is evaluated as
\begin{eqnarray}
T_{1\rightarrow 0}&\approx&\int_{x^{\ast}}^0 \frac{1}{ D(y)} \exp (-\Phi(y))dy\int_{1} ^ {x_{0}^{\ast}} \exp(\Phi(z))dz \nonumber \\
&\approx&\frac{ 2\pi\alpha'\beta'\exp(\Phi(x_{0}^{\ast})-\Phi(x^{\ast}))}{D(x^{\ast})}\nonumber\\
&\propto&\frac{1}{D(x^{\ast})}\exp(\Phi(x_{0}^{\ast})-\Phi(x^{\ast})).
\end{eqnarray}
From the expression of Eq.(44), the single click time is not sensitive to the assumption of Eq.(40).\\
\indent In the higher mutation rates regime,  where $x_{0}^{\ast}$ approximates to an adaptive state which is near enough to $1$, $x^{\ast}$ corresponds to the unadaptive state that the population lies between the adaptive states $0$ and $x_{0}^{\ast}$.  The potential barrier is finite. According to classical derivation corresponding to Eq.(44) the single click time approximates to
\begin{eqnarray}
T_{1\rightarrow 0}&&= \lim_{x\rightarrow 1}\int_{0}^{x} \dfrac{1}{\epsilon D(y)}\exp (-\Phi(y))dy \int_{y}^ 1  \exp(\Phi(z))dz  \nonumber  \\
~&&\approx 2N\int_{0}^{x^{\ast}}\dfrac{(1-y)^{2N\mu(\sigma-1)/(1-\sigma\mu)}}{(1-\sigma+y\sigma(1-\mu))^{2N(1-\mu)/(1-\sigma\mu)}}dy\nonumber\\
&& \quad\times\int_{x_{0}^{\ast}}^{1}z^{-1}(1-z)^{(\sigma\mu-1-2N\mu(\sigma-1))/(1-\sigma\mu)}\nonumber\\
&& \quad \times(1-\sigma+\sigma z(1-\mu))^{2N(1-\mu)/(1-\sigma\mu)}dz\nonumber\\
~~&&\approx\frac{1}{D(x^{\ast})}\exp(\Phi(x_{0}^{\ast})-\Phi(x^{\ast}))\nonumber\\
&&\approx  \frac{N(N-1)^{2}\sigma^{2}(1-\mu)^{2}}{(\alpha-\sqrt{\alpha^{2}-\beta})(4\sigma(N-1)(1-\mu)-\alpha+\sqrt{\alpha^{2}-\beta})}.
\end{eqnarray}
Here $x_{0}^{\ast}$ is the fixed point $x_{2}$, and $x^{\ast}$ is the fixed point $x_{1}$, parameters $\alpha$, $\beta$ is the same as Eqs.(34) (35) respectively. The difference of potential is
\begin{eqnarray}
\Phi(x_{0}^{\ast})-\Phi(x^{\ast})&=&\frac{2N\mu(1-\sigma)-1+\sigma\mu}{1-\sigma\mu}\ln \left(1-\frac{x_{2}-x_{1}}{1-x_{1}}\right)-\ln\left(1+\frac{x_{2}-x_{1}}{x_{1}}\right)\nonumber\\
&&+\frac{2N(1-\mu)}{1-\sigma\mu}\ln\left(1+\frac{\sigma(1-\mu)(x_{2}-x_{1})}{1-\sigma+x_{1}\sigma(1-\mu)}\right)\nonumber\\
&=&\frac{2N\mu(1-\sigma)}{1-\sigma\mu}\ln\left(1-\frac{2\sqrt{\alpha^{2}-\beta}}{\alpha-\sqrt{\alpha^{2}-\beta}}\right)\nonumber\\
&&-\ln\left(1-\frac{4(\alpha^{2}-\beta)}{(\alpha-\sqrt{\alpha^{2}-\beta})^{2}}\right)\nonumber\\
&&+\frac{2N(1-\mu)}{1-\sigma\mu}\ln\left(1+\frac{2\sqrt{\alpha^{2}-\beta}}{4N-6-6N\sigma+7\sigma-\sigma\mu+2N\mu+\sqrt{\alpha^{2}-\beta}}\right),
\end{eqnarray}
Where $\alpha$ and $\beta$ are the same as Eqs.(34) and (35)The approximated single click time varies with mutation rates in Fig.4. The single click time $T_{1\rightarrow 0}$ increases with population size $N$ in certain regime, decreases with mutation rates $\mu$ and selection rates $\sigma$ in the parameters regime $\mu\in(2N/4N(N-1),1)$ and $\sigma\in((2+2\mu-10N\mu+4N^{2}\mu+2N\mu^{2}+4(1-\mu)\sqrt{N(N-1)((2N-1)\mu-1)})/(\mu-2N+1)^{2},(2N\mu-1)/(2N\mu-\mu))$.  Because in the regime, with selection rates increasing, the difference of potential between two fixed points decreases as the demonstration in Fig.2, the viability of suboptimal class decreases, populations evolve to the fittest class. These results in the the single click time shorter. In another hand, with deleterious mutation increasing, the population of suboptimal class increases, the difference of potential between two fixed points decreases as the demonstration in Fig.2. These also results in the single click time shorter. \\
\indent For the lower mutation rates regime, where the potential barrier is infinite. The single click time can be estimated also,
$x^{\ast}$ corresponds to the fixed point $x_{1}$ that the population lies at the lowest potential.
\begin{eqnarray}
  T_{1\rightarrow 0}
~&&\approx 2N\int_{0}^{1}\dfrac{(1-y)^{2N\mu(\sigma-1)/(1-\sigma\mu)}}{(1-\sigma+y\sigma(1-\mu))^{2N(1-\mu)/(1-\sigma\mu)}}dy\nonumber\\
&& \quad\times\int_{x_{0}^{\ast}}^{1}z^{-1}(1-z)^{(\sigma\mu-1-2N\mu(\sigma-1))/(1-\sigma\mu)}\nonumber\\
&& \quad \times(1-\sigma+\sigma z(1-\mu))^{2N(1-\mu)/(1-\sigma\mu)}dz\nonumber\\
&&\approx \frac{1-\sigma\mu}{\mu(1-\sigma)}.
\end{eqnarray}
 From expression of Eq.(47), the single click time goes to infinity with mutation rates tends to zero in the parameters regimes of $\mu\in(0,1/(2N-1))$ and $\sigma\in(\mu,1)$. And when parameters regions lie $\mu\in(1/(2N-1),1)$ and $\sigma\in((2N\mu-1)/(2N\mu-\mu),1)$, the results of the single click time is not sensitive to the population size. Biologically if deleterious mutation accumulates, the viability of suboptimal class increases as the demonstration in Fig.2, these results in the single click time longer.\\
\indent Analogous to the derivation of $T_{1\rightarrow 0}$, we can calculate
\begin{eqnarray}
  \lefteqn{T_{0\rightarrow 1}}~
&&= \lim_{x\rightarrow 0}\int_{1}^{x} \dfrac{1}{\epsilon D(y)}\exp (-\Phi(y))dy \int_{y}^ 0  \exp(\Phi(z))dz  \nonumber  \\
&&\approx \lim_{x\rightarrow 0}2N\int_{1}^{x}\dfrac{(1-y)^{2N\mu(\sigma-1)/(1-\sigma\mu)}}{(1-\sigma+y\sigma(1-\mu))^{2N(1-\mu)/(1-\sigma\mu)}}dy\nonumber\\
&& \quad \times\int_{y}^{0}z^{-1}(1-z)^{(\sigma\mu-1-2N\mu(\sigma-1))/(1-\sigma\mu)}\nonumber\\
&& \quad \times(1-\sigma+\sigma z(1-\mu))^{2N(1-\mu)/(1-\sigma\mu)}dz\nonumber \\
&&\approx2N\int_{0}^{1}\dfrac{(1-y)^{2N\mu(\sigma-1)/(1-\sigma\mu)}}{(1-\sigma+y\sigma(1-\mu))^{2N(1-\mu)/(1-\sigma\mu)}}dy\nonumber\\
&&\quad\times\int_{0}^{y}(z^{-1}+\ldots)dz\nonumber\\
&& \rightarrow\infty.
\end{eqnarray}
\indent Compared with the singular point $x=1$, the difference between two singular points $x=0$ and $x=1$ is the mutation rates $\mu$. The mutation rates $\mu>0$ at $x=1$, this results in the power of $(1-z)$ is not negative, so the single click time is finite. The infinity of the single click time from $x=0$ comes from that the mutation from unfavored allele $a$ to favored allele $A$ is zero, this results in the second integral nonintegrable because of the negative power of $z$. These results are consistent with biological inference. Biologically because the absence of back mutation, the accumulation of deleterious mutation, once the population arrives at the state that almost individuals are with allele $a$, the population is absorbed to the state and can not leave with high probability. Mathematically, because the second integral is singular  for the singular point $x=0$. And the integrated function is a fraction respect to argument $x$, but the highest power of denominator is smaller than that of numerator. That results in the power of $x$ is $2N-2$. As a consequence the second integral is singular.
\section*{Discussion}
\indent We analytically construct adaptive landscape. The constructive method is independent on the existence and normalization of stationary distribution. We demonstrate the position and adaptiveness of all fixed points for the whole parameters regimes under the condition of the diffusion approximation. An interesting thing is the imaginary fixed points occurring. We give the parameters regions of their occurrence. However, we have not found any study of Muller ratchet for the fixed points to give a complete description. In addition, we give the description of escape from infinite potential. However, intuitively infinite potential means the population lies at adaptive state. The transition from the adaptive state can not occur. Here we find that the escape from infinite potential can not occur when the boundary is absorbing.  So we define the absorbing boundary by adaptive landscape and the single click time without any extraneous assumptions.  \\
\indent The model with discrete manner describes the nature of populations evolution. Here we give two special cases. One is that the population lies at that state with  allele frequency $x=0$, the other is that the population lies at the state with  allele frequency $x=1$. We compare the model with discrete and continuous manners to conclude the definition of boundary conditions without any extraneous assumptions.  The model with continuous manner is derived under the condition of enough small space change $\delta x$ in one generation, in another word, when population size $N$ is much bigger. The model with continuous manner can correspond to the model with discrete manner.\\
  \indent When allele frequency $x=0$, the matrix for transition probabilities  demonstrates that the population only absorbs the flux from the next, and it can not output any flux. In the model with continuous manner the value of adaptive landscape with  allele frequency $x=0$ arrives at the maximum except $\sigma=1$. This demonstrates the state with allele frequency $x=0$ is almost always adaptive. Then we calculate the single click time from the  boundary with allele frequency $x=0$. We find the single click time is infinite. So we draw the conclusion that the state with allele frequency $x=0$  absorbs flux, and does not output flux. The boundary is absorbing. When the allele frequency $x=1$ in the model with discrete manner, the transition probability is 
\begin{equation} (N!(\mu(1-\sigma))^{N-m}(1-\mu)^{m})/(m!(N-m)!(1-\sigma\mu)^{N}),  \nonumber
\end{equation}
that is, when $\mu> 0$, the transition probabilities are not zero, the state with allele frequency $x=1$ can input any flux to its next state. The single click time from the state is finite in the model with diffusion manner, however the potential at the state could be infinite, then we draw the conclusion  this boundary is not absorbing. This is consistent with the biological understanding. \\
\indent This article presents an approach to estimate the single click time of Muller's ratchet. Furthermore, it define the absorbing phenomenon by the single click time without any extraneous assumptions.  Inspired by \cite{Higgs1996} \cite{waxman2010}, we connect Muller's ratchet to one locus Wright-Fisher model with asexual population including $N$ haploid individuals. And our model is represented as a Fokker-Planck equation. We give a complete description for the position and adaptiveness of all fixed points in the whole parameters regimes. This is first done bases on  diffusion approximation. The investigated elements are at the allele level. This is different from Ref. \cite{waxman2010}.
  Our method does not need the existence and normalization of the stationary distribution. Our constructive method is independent of the stationary distribution.  Compared with the method based on  diffusion approximation \cite{Gessler1995}\cite{Etheridge2009}, mathematically it is described as stochastic differential equations. Our method investigates the global dynamical property of the system, and reduces the complexity of calculating  stochastic differential equations. In addition, the boundary condition of these stochastic differential equations is prescribed.  Compared with Ref. \cite{waxman2010}, They added Dirac function to the boundary. But this is not appropriate for the adding non-differential Dirac function to stationary distribution, and stationary distribution should satisfy diffusion equation. However, the treatment is convenient for computing the stationary distribution. The stationary distribution of theirs is equivalent to our adaptive landscape. They had not given the shape of adaptive landscape when the mutation rate lies in the lower regime.   We use the model defined in the interval $(0,1)$ to describe the  absorbing boundary. We check the biological phenomenon  by  the model with both discrete and continuous manners.  This is a new method to handle the boundary condition. We investigate the absorbing phenomenon by it without any extraneous assumptions.   \\
  \indent To summarize, we have obtained two main sets of results in the
present work.    Most importantly, we find that  the absorbing phenomenon  is characterized by the adaptive landscape and the single click time without any extraneous assumptions.  First, we demonstrate the adaptive landscape can be
explicitly read out as a potential function from general diffusion equation.
This not only allows computing the single click time of Muller's ratchet
straightforward, but also characterizes the whole picture of the ratchet
mechanism. The adaptive landscape has rich structures such as finite  and infinite potential, real  and imaginary fixed points. We  analytically demonstrate the adaptive and unadaptive states for the whole parameters regimes. We find corresponding parameters regimes for different shapes of adaptive landscape.  Second, we give the formula about the single click time with finite and infinite potential. And we find the single click time with finite potential increases with  selection rates and population size increasing, decreases with mutation rates increasing, the single click time with infinite potential is insensitive to the population size. These results give a new understanding of infinite potential and allow us a new way to handle the absorbing
phenomenon. In this perspective our work may be a starting
point for estimating the click time for Muller's ratchet in more
general situations and for describing the boundary condition. Such demonstration suggests that adaptive landscape may be applicable to other levels of systems biology.
\bigskip

\section*{Author's contributions}
   \indent  Shuyun Jiao carried the research and writing.  Ping Ao  oversaw the whole project, and participated in research and writing.

\section*{Acknowledgements}
  \ifthenelse{\boolean{publ}}{\small}{}
   \indent We would like to thank  Yanbo Wang for drawing  the figures, also thank Quan Liu for discussions and technical help, thank Song Xu for technical help. We thank Bo Yuan for some advice on writing and correcting some expression on language. This work was supported in part by the National 973
Projects No. 2010CB529200
(P.A.), and in part by No. 91029738 (P.A.) and No.Z-XT-003(S.J.) and by the project of Xinyang Normal university No. 20100073(S.J.).
\section*{Competing interests}
 \indent The authors declare that they have no competing financial interests.


{\ifthenelse{\boolean{publ}}{\footnotesize}{\small}
 \bibliographystyle{bmc_article}  
  \bibliography{bmc_article4} }     
\begin{figure}[h]
\centering
\includegraphics[width=12cm]{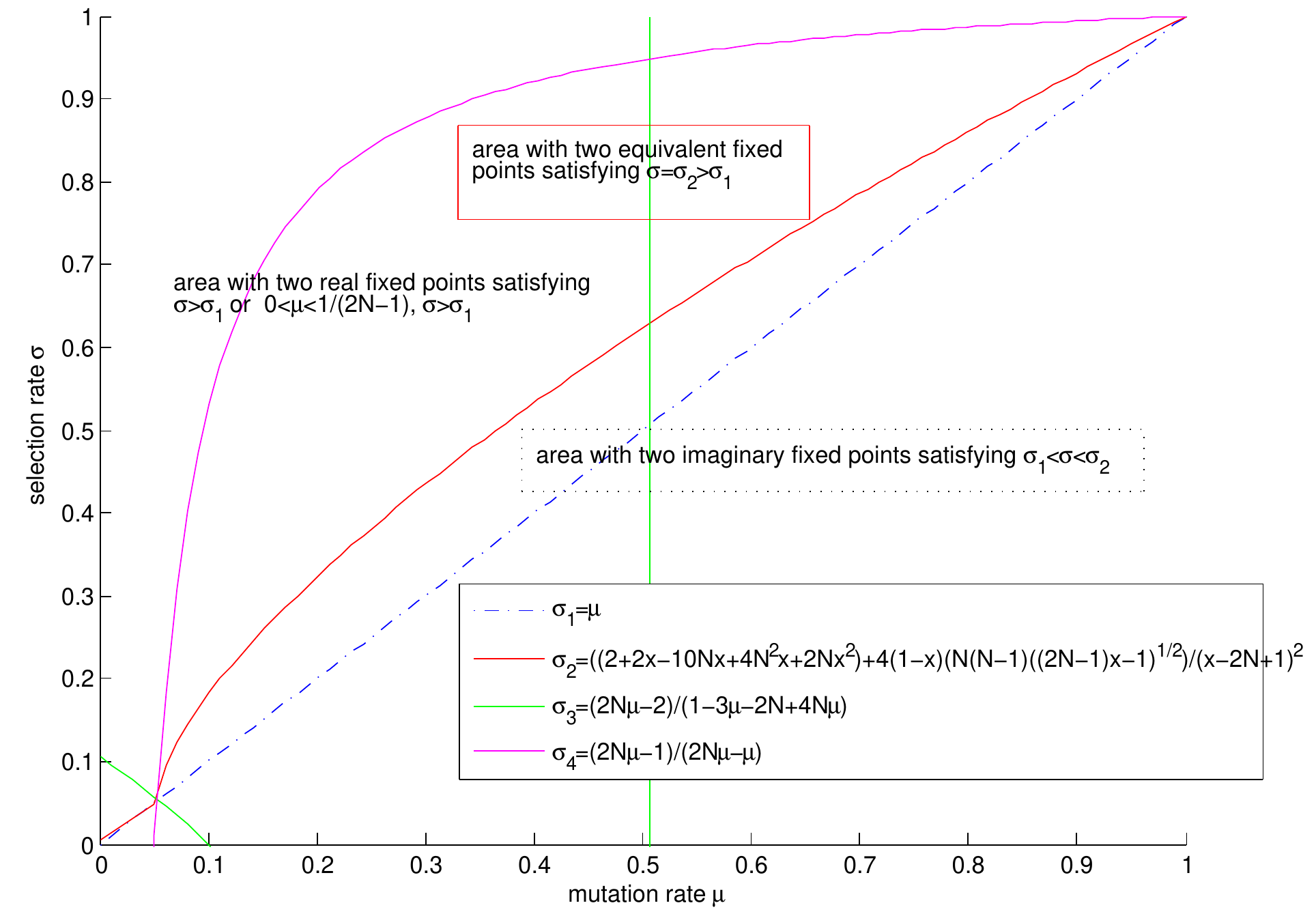}
\caption{\small \indent  Relation of fixed points and parameters for the system in all regimes. The regime represented by III i) with parameters regions $\sigma\in(\sigma_{1},\sigma_{2})$ and $\mu\in(2/(2N-1+2\sqrt{N(N-1)}),1)$ has two imaginary fixed points. The red curve denoted by II) with parameters satisfying $\sigma=\sigma_{2}$ and $\mu\in(2/(2N-1+2\sqrt{N(N-1)}),1)$  has  two equivalent fixed points. The three cases of II)  occur in the intervals. The regimes denoted by I) with parameters satisfying $\sigma\in(\sigma_{2},\sigma_{5})$ and $\mu\in(2/(2N-1+2\sqrt{N(N-1)}),(2N-1)/4N(N-1))$; satisfying $\sigma\in(\sigma_{2},\sigma_{5})$ and $\mu\in((2N-1)/4N(N-1),1)$; satisfying $\sigma\in(\sigma_{1},\sigma_{5})$ and  $\mu\in(0,1/(2N-1))$; satisfying $\sigma\in(\sigma_{1},\sigma_{5})$ and $\mu\in(1/(2N-1),2/(2N-1+2\sqrt{N(N-1)}))$ have two real fixed points. The five cases of I) occur in the regions.   }
\end{figure}
\begin{figure}[h]
\centering
\includegraphics[width=12cm]{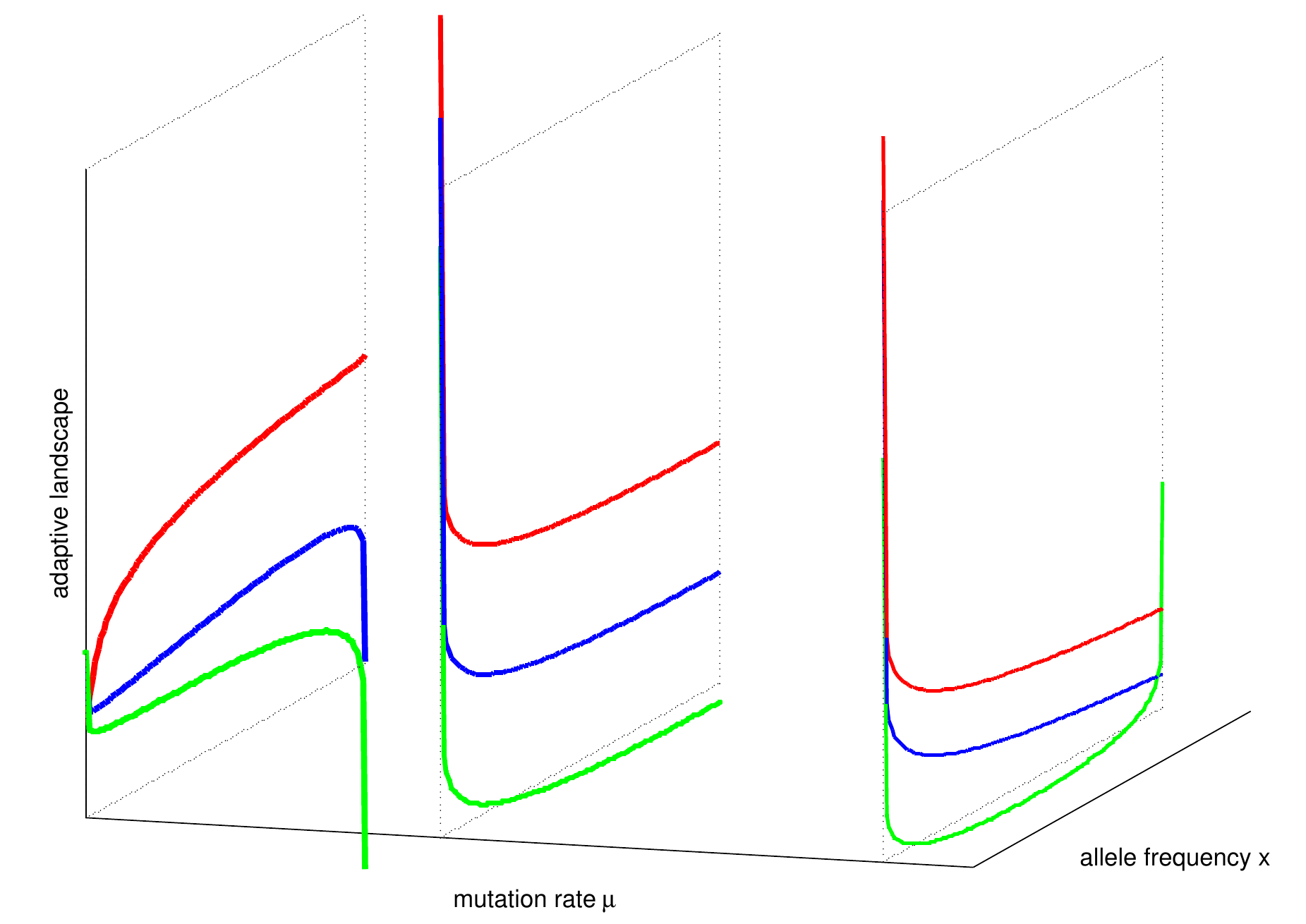}
\caption{\small \indent Adaptive landscape against allele frequency $x$ with mutation rates and selection rates in all regimes. $x$ label represents allele frequency, $\mu$ label represents mutation rates while vertical label is corresponding value of $\Phi$.  Assume population size is constant $N$=100. In the right graph section, green line corresponds to the case of I iii), where parameters satisfy $\mu$=0.00005, $\sigma$=0.0005. Blue line approximates to the case of III, where parameters satisfy $\mu$=1/198, $\sigma$=1/198. Red line corresponds to the case of I i), where parameters satisfy $\mu$=0.0050252, $\sigma$=0.00505. In the middle graph section, green line represents the case of I  ii), where parameters satisfy $\mu$=0.0050252, $\sigma$=0.005039925. Blue line represents the case of II ii), where parameters satisfy $\mu$=0.0050252, $\sigma$=0.0050503775.  Red line represents the case of II i), where parameters satisfy $\mu$=0.005025, $\sigma$=0.00503. In the left graph section, green line represents the case of, where parameters satisfy $\mu$=0.04, $\sigma$=0.06 II iii). Blue line represents the case of I vi), where parameters satisfy $\mu$=0.04, $\sigma$=0.19.  Red line represents the case of I iv), where parameters satisfy $\mu$=0.04, $\sigma$=0.99.}
\end{figure}
\begin{figure}[h]
\centering
\includegraphics[width=12cm]{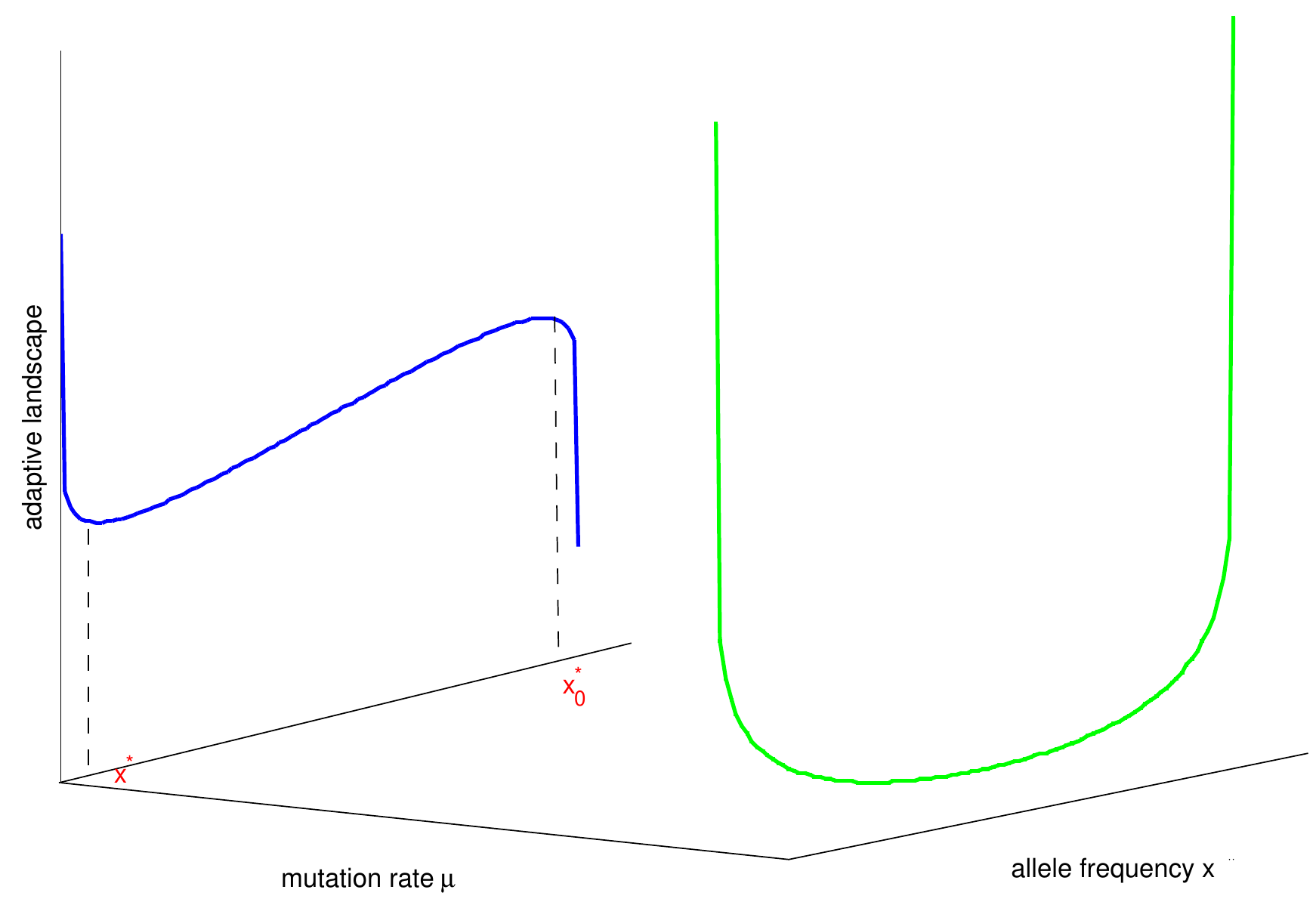}
\caption{\small \indent Adaptive landscape with two adaptive states. $N$=100, blue line  represents $\mu$=0.04,$\sigma$=0.19 corresponding the case for I vi) while green line stands for $\mu=0.000005$,  $\sigma$=0.00005 corresponding the case for I iii).}
\end{figure}
\begin{figure}[h]
\centering
\includegraphics[width=12cm]{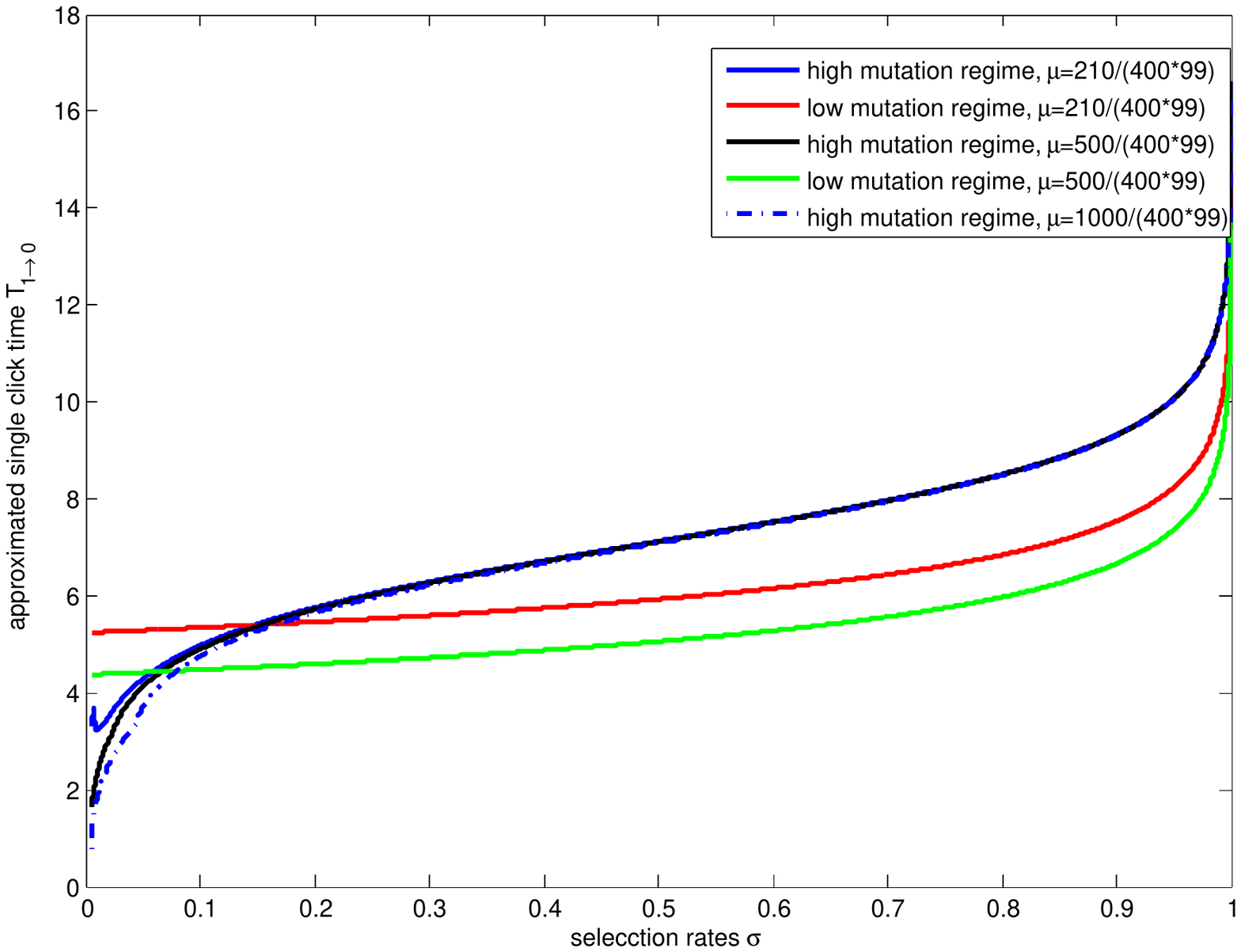}
\caption{\small \indent The approximated single click time decreases with mutation rates increasing and selection rates decreasing in the regime $\mu\in(2N/4N(N-1),1)$ and $\sigma\in((2+2\mu-10N\mu+4N^{2}\mu+2N\mu^{2}+4(1-\mu)\sqrt{N(N-1)((2N-1)\mu-1)})/(\mu-2N+1)^{2},(2N\mu-1)/(2N\mu-\mu))$ denoted by I iv). Assume $N$=100, blue, black and dashed lines, corresponding high mutation rates regime, represent  $\mu=210/(400*99)$, $\mu=500/(400*99)$ and $\mu=1000/(400*99)$ respectively. Red and green lines, corresponding low mutation rates regime, represent $\mu=210/(400*99)$ and $\mu=500/(400*99)$ respectively. The approximated single click time increases with mutation rates increasing in the regime $\mu\in(1/(2N-1),1)$ and $\sigma\in((2N\mu-1)/(2N\mu-\mu),1)$ denoted by I iii). Green dotline represents this case.}
\end{figure}

\ifthenelse{\boolean{publ}}{\end{multicols}}{}

\end{document}